\documentclass[10pt,a4paper]{article}
\usepackage{amsmath}
\usepackage{amssymb}
\usepackage{amsfonts}
\usepackage{amstext}
\usepackage{amsbsy}
\usepackage[mathscr]{eucal}
\usepackage{graphicx}
\usepackage{color}
\usepackage[all]{xy}
\newcommand{\beq}{\begin{equation}}
\newcommand{\eeq}{\end{equation}}
\newcommand{\beqa}{\begin{eqnarray}}
\newcommand{\eeqa}{\end{eqnarray}}
\newcommand{\ket}[1]{| #1 \rangle}
\newcommand{\bra}[1]{\langle #1 |}

\title{\Large\textbf{Geometrical structures of multipartite quantum systems}}

\author{\textit{ Hoshang Heydari}\\
        \small\textit{Physics Department, Stockholm university 10691 Stockholm Sweden}\\
\\\small\textit{Email: hoshang@fysik.su.se}}
\begin{document}
\maketitle

\begin{abstract}
 In this paper I will investigate geometrical structures of multipartite quantum systems based on complex projective varieties. These varieties are important in characterization of quantum entangled states. In particular I will establish relation between multi-projective Segre varieties and multip-qubit quantum states. I also will discuss other geometrical approaches such as toric varieties to visualize complex multipartite quantum systems.
\end{abstract}


\maketitle
\section{Introduction}
Characterization of multipartite quantum systems is very interesting research topic in the foundations of quantum theory and has many applications in the field quantum information and quantum computing.
And in particular geometrical structures of multipartite quantum entangled
pure states are of special importance.
 In this paper we will review the construction of
  Segre variety for multi-qubit states. We will also show a
   construction of geometrical measure of entanglement based on
    the Segre variety for multi-qubit systems. Finally we will
    establish a relation between the Segre variety, toric variety,
    and multi-qubit quantum systems. The relation could be used as a
    tool to visualize entanglement properties of multi-qubit states.
Let  $\mathcal{Q}_{j},~j=1,2,\ldots m$ be quantum systems  with underlying Hilbert spaces $\mathcal{H}_{\mathcal{Q}_{j}}$. Then the Hilbert space of a multi-qubit systems $\mathcal{Q}$, is given by $\mathcal{H}_{\mathcal{Q}}=\mathcal{H}_{\mathcal{Q}_{m}}\otimes \mathcal{H}_{\mathcal{Q}_{m-1}}\otimes\cdots\otimes \mathcal{H}_{\mathcal{Q}_{1}}$, where $\mathcal{H}_{\mathcal{Q}_{j}}=\mathbf{C}^{2}$ and $\dim \mathcal{H}_{\mathcal{Q}}=2^{m}$. Now, let
\begin{equation}\ket{\Psi}=\sum^{1}_{x_{m}=0}\sum^{1}_{x_{m-1}=0}\cdots
\sum^{1}_{
x_{1}=0}\alpha_{x_{m}x_{m-1}\cdots x_{1}}\ket{x_{m}x_{m-1}\cdots
x_{1}},
\end{equation}
be a vector in $\mathcal{H}_{\mathcal{Q}}$,
where
$\ket{x_{m}x_{m-1}\cdots~x_{1}}=
\ket{x_{m}}\otimes\ket{x_{m-1}}\otimes\cdots\otimes\ket{x_{1}}$ are orthonormal basis in
$\mathcal{H}_{\mathcal{Q}}$
 and $\alpha_{x_{m}x_{m-1}\cdots x_{1}}\in \mathbf{C}$. Then the  quantum states are normalized vectors in $\mathcal{P}(\mathcal{H}_{\mathcal{Q}})= \mathcal{H}_{\mathcal{Q}}/\sim$. Moreover, let
$\rho_{\mathcal{Q}}=\sum^{\mathrm{N}}_{i=1}p_{i}\ket{\Psi_{i}}\bra{\Psi_{i}}$,
for all $0\leq p_{i}\leq 1$ and $\sum^{\mathrm{N}}_{i=1}p_{i}=1$,
denotes a density operator acting on the Hilbert space $\mathcal{H}_{\mathcal{Q}}$.

The density operator
$\rho_{\mathcal{Q}}$ is said to be fully separable, which we will
denote by $\rho^{sep}_{\mathcal{Q}}$, with respect to the Hilbert
space decomposition, if it can  be written as $
\rho^{sep}_{\mathcal{Q}}=\sum^\mathrm{N}_{i=1}p_i
\bigotimes^m_{j=1}\rho^i_{\mathcal{Q}_{j}},
$ where  $\rho^i_{\mathcal{Q}_{j}}$ denotes a density operator on
Hilbert space $\mathcal{H}_{\mathcal{Q}_{j}}$. If
$\rho^{p}_{\mathcal{Q}}$ represents a pure state, then the quantum
system is fully separable if $\rho^{p}_{\mathcal{Q}}$ can be written
as
$\rho^{sep}_{\mathcal{Q}}=\bigotimes^m_{j=1}\rho_{\mathcal{Q}_{j}}$,
where $\rho_{\mathcal{Q}_{j}}$ is the density operator on
$\mathcal{H}_{\mathcal{Q}_{j}}$. If a state is not separable, then
it is said to be an entangled state.

\section{Projective geometry}

In this section we give a short introduction to variety.
Let $\mathbf{C}[z]=\mathbf{C}[z_{1},z_{2}, \ldots,z_{n}]$ denotes the polynomial
algebra in $n$  variables with complex coefficients. Then, given a
set of $r$ polynomials $\{g_{1},g_{2},\ldots,g_{r}\}$ with $g_{i}\in
\mathbf{C}[z]$, we define a complex affine variety as
\begin{eqnarray}
&&\mathcal{V}_{\mathbf{C}}(g_{1},g_{2},\ldots,g_{r})=\{P\in\mathbf{C}^{n}:
g_{i}(P)=0~\forall~1\leq i\leq r\},
\end{eqnarray}
where $P\in\mathbf{C}^{n}$ is called a point of $\mathbf{C}^{n}$ and if
$P=(a_{1},a_{2},\ldots,a_{n})$ with $a_{j}\in\mathbf{C}$, then
$a_{j}$ is called the coordinates of $P$.
A complex projective space $\mathbf{CP}^{n}$ is defined to be the
set of lines through the origin in $\mathbf{C}^{n+1}$, that is,
\begin{equation}
\mathbf{CP}^{n}=\frac{\mathbf{C}^{n+1}-\{0\}}{
u\sim v},~\lambda\in
\mathbf{C}-0,~v_{i}=\lambda u_{i} ~\forall ~0\leq i\leq n+1,
\end{equation}
where $u=(u_{1},\ldots,u_{n+1})$ and $v=(v_{1},\ldots,v_{n+1})$.  Given a set of homogeneous polynomials
$\{g_{1},g_{2},\ldots,g_{r}\}$  with $g_{i}\in \mathbf{C}[z]$, we define a
complex projective variety as
\begin{eqnarray}
&&\mathcal{V}(g_{1},\ldots,g_{r})=\{O\in\mathbf{CP}^{n}:
g_{i}(O)=0~\forall~1\leq i\leq r\},
\end{eqnarray}
where $O=[a_{1},a_{2},\ldots,a_{n+1}]$ denotes the equivalent class
of point $\{\alpha_{1},\alpha_{2},\ldots,$
$\alpha_{n+1}\}\in\mathbf{C}^{n+1}$. We can view the affine complex
variety
$\mathcal{V}_{\mathbf{C}}(g_{1},g_{2},\ldots,g_{r})\subset\mathbf{C}^{n+1}$
as a complex cone over the complex projective variety
$\mathcal{V}(g_{1},g_{2},\ldots,g_{r})$.

  We can map the
product of  spaces $\underbrace{\mathbf{CP}^{1}\times\mathbf{CP}^{1}
\times\cdots\times\mathbf{CP}^{1}}_{m~\mathrm{times }}$ into a projective space by
its Segre embedding as follows.  The Segre map is given by
\begin{equation}
\begin{array}{ccc}
  \mathcal{S}_{2,\ldots,2}:\mathbf{CP}^{1}\times\mathbf{CP}^{1}
\times\cdots\times\mathbf{CP}^{1}&\longrightarrow&
\mathbf{CP}^{2^{m}-1},\\
\end{array}
\end{equation}
is defined by $ ((\alpha^{1}_{0},\alpha^{1}_{1}),\ldots,
 (\alpha^{m}_{0},\alpha^{m}_{1}))  \longmapsto
 (\alpha^{m}_{i_{m}}\alpha^{m-1}_{i_{m-1}}\cdots\alpha^{1}_{i_{1}})$, where $(\alpha^{i}_{0},\alpha^{i}_{1})$  is
points defined on the $i$th complex projective space
$\mathbf{CP}^{1}$ and $\alpha_{i_{m}i_{m-1}\cdots i_{1}}$,$0\leq i_{s}\leq 1$
be a homogeneous coordinate-function on
$\mathbf{CP}^{2^{m}-1}$. Moreover, let us consider
a multi-qubit quantum system
 and let
$
\mathcal{A}=\left(\alpha_{i_{m}i_{m-1}\ldots i_{1}}\right)_{0\leq
i_{s}\leq 1},
$
for all $j=1,2,\ldots,m$. $\mathcal{A}$ can be realized as the
following set $\{(i_{1},i_{2},\ldots,i_{m}):1\leq i_{s}\leq
2,\forall~s\}$, in which each point $(i_{m},i_{m-1},\ldots,i_{1})$
is assigned the value $\alpha_{i_{m}i_{m-1}\ldots i_{1}}$.  For each $s=1,2,\ldots,m$, a two-by-two minor about the
$j$-th coordinate of $\mathcal{A}$ is given by
\begin{eqnarray}\label{segreply1}
&&\mathcal{I}^{m}_{\mathcal{A}}=
\alpha_{x_{m}x_{m-1}\ldots x_{1}}\alpha_{y_{m}y_{m-1}\ldots y_{1}}
-
\alpha_{x_{m}x_{m-1}\ldots x_{s+1}y_{s}x_{s-1}\ldots
x_{1}}\alpha_{y_{m}y_{m-1} \ldots y_{s+1} x_{s}y_{s-1}\ldots y_{m}}.
\end{eqnarray}
Then the ideal $\mathcal{I}^{m}_{\mathcal{A}}$ is generated by
The image of the Segre embedding
$\mathrm{Im}(\mathcal{S}_{2,2,\ldots,2})$, which again
is an intersection of families of quadric hypersurfaces in
$\mathbf{CP}^{2^{m}-1}$, is called Segre variety
and it is given by
\begin{eqnarray}\label{eq: submeasure}
\mathrm{Im}(\mathcal{S}_{2,2,\ldots,2})&=&\bigcap_{\forall
s}\mathcal{V}\left(\mathcal{I}^{m}_{\mathcal{A}}\right).
\end{eqnarray}
This is the space of separable multi-qubit states.
Moreover, we propose a measure of
entanglement for general pure multipartite states based on modified Segre variety  as follows
\begin{eqnarray}\label{EntangSeg2}\nonumber
&&\mathcal{F}(\mathcal{Q}^{p}_{m}(2,2\ldots,2))
=(\mathcal{N}\sum_{\forall \sigma\in\mathrm{Perm}(u)}\sum_{
k_{j},l_{j}, j=1,2,\ldots,m}\\&&|\alpha_{k_{1}k_{2}\ldots
k_{m}}\alpha_{l_{1}l_{2}\ldots l_{m}} -
\alpha_{\sigma(k_{1})\sigma(k_{2})\ldots\sigma(k_{m})}\alpha_{\sigma(l_{1})\sigma(l_{2})
\ldots\sigma(l_{m})}|^{2})^{\frac{1}{2}},
\end{eqnarray}
where $\sigma\in\mathrm{Perm}(u)$ denotes all possible sets of
permutations of indices for which $k_{1}k_{2}\ldots k_{m}$ are
replace by $l_{1}l_{2}\ldots l_{m}$, and $u$ is the number of
indices to permute.

As an example we will discuss the four-qubit state
in which we first encounter
these new varieties. For this quantum system we can partition the
Segre embedding as follows:
$$\xymatrix{ \mathbf{P}^{1}\times\mathbf{P}^{1}\times\mathbf{P}^{1}\times\mathbf{P}^{1}
\ar[d]_{\mathcal{S}_{2,\ldots,2}}\ar[r]_{\mathcal{S}_{2,2}\otimes
I\otimes I}&\mathbf{P}^{3}
\times\mathbf{P}^{1}\times\mathbf{P}^{1}\ar[d]_{I\otimes\mathcal{S}_{2,2}}\\
             \mathbf{P}^{2^{4}-1}&\mathbf{P}^{3}
             \times\mathbf{P}^{3}\ar[l]_{\mathcal{S}_{4,4}}}.$$
 For the Segre variety,
which is represented by completely decomposable tensor, we have a
commuting diagram and
$\mathcal{S}_{2,\ldots,2}=(\mathcal{S}_{4,4})
\circ(I\otimes\mathcal{S}_{2,2})\circ(\mathcal{S}_{2,2}\otimes
I\otimes I)$.
\section{Toric variety and multi-qubit quantum systems}

Let $S\subset \mathbf{R}^{n}$ be finite subset, then a convex polyhedral cone is defined by
$
 \sigma=\mathrm{Cone}(S
 )=\left\{\sum_{v\in S}\lambda_{v}v|\lambda_{v}\geq0\right\}.$
In this case $\sigma$ is generated by $S$.  In a similar way we define  a polytope by
$
 P=\mathrm{Conv}(S)=\left\{\sum_{v\in S}\lambda_{v}v|\lambda_{v}\geq0, \sum_{v\in S}\lambda_{v}=1\right\}.
$
We also could say that $P$ is convex hull of $S$. A convex polyhedral cone is called simplicial if it is generated by linearly independent set. Now, let $\sigma\subset \mathbf{R}^{n}$ be a convex polyhedral cone and $\langle u,v\rangle$ be a natural pairing between $u\in \mathbf{R}^{n}$ and $v\in\mathbf{R}^{n}$. Then, the dual cone of the $\sigma$ is define by
$$
 \sigma^{\wedge}=\left\{u\in \mathbf{R}^{n*}|\langle u,v\rangle\geq0~\forall~v\in\sigma\right\},
$$
where $\mathbf{R}^{n*}$ is dual of $\mathbf{R}^{n}$.
We call a convex polyhedral cone strongly convex if $\sigma\cap(-\sigma)=\{0\}$.

The algebra of Laurent polynomials is defined by
$
\mathbf{C}[z,z^{-1}]=\mathbf{C}[z_{1},z^{-1}_{1},\ldots,z_{n},z^{-1}_{n}],
$
where $z_{i}=\chi^{e^{*}_{i}}$. The terms  of the form $\lambda \cdot z^{\beta}=\lambda z^{\beta_{1}}_{1}z^{\beta_{2}}_{2}\cdots z^{\beta_{n}}_{n}$ for $\beta=(\beta_{1},\beta_{2},\ldots,\beta_{n})\in \mathbf{Z}$ and $\lambda\in \mathbf{C}^{*}$ are called Laurent monomials. A ring $R$ of Laurent polynomials is called a monomial algebra if it is a $\mathbf{C}$-algebra generated by Laurent monomials. Moreover, for a lattice cone $\sigma$, the ring
$$R_{\sigma}=\{f\in \mathbf{C}[z,z^{-1}]:\mathrm{supp}(f)\subset \sigma\}
$$
is a finitely generated monomial algebra, where the support of a Laurent polynomial $f=\sum_{i}\lambda_{i}z^{i}$ is defined by
$$\mathrm{supp}(f)=\{i\in \mathbf
{Z}^{n}:\lambda_{i}\neq0\}.$$
 Now, for a lattice cone $\sigma$ we can define an affine toric variety to be the maximal spectrum $$\mathbf{X}_{\sigma}=\mathrm{Spec}R_{\sigma}.$$  A toric variety
$\mathbf{X}_{\Sigma}$ associated to a fan $\Sigma$ is the result of gluing affine varieties
$\mathbf{X}_{\sigma}=\mathrm{Spec}R_{\sigma}$ for all $\sigma\in \Sigma$  by identifying $\mathbf{X}_{\sigma}$ with the corresponding Zariski open subset in $\mathbf{X}_{\sigma^{'}}$ if
$\sigma$ is a face of $\sigma^{'}$. That is,
first we take the disjoint union of all affine toric varieties $\mathbf{X}_{\sigma}$ corresponding to the cones of $\Sigma$.
Then by gluing all these affine toric varieties together we get $\mathbf{X}_{\Sigma}$.

A compact toric variety $\mathcal{X}_{A}$ is called projective if there exists an injective morphism
$$\Phi:\mathcal{X}_{\Sigma}\longrightarrow\mathbf{P}^{r}$$ of $\mathcal{X}_{\Sigma}$
 into some projective space such that $\Phi(\mathcal{X}_{\Sigma})$ is Zariski
 closed in $\mathbf{P}^{r}$.
A toric variety $\mathcal{X}_{\Sigma}$ is equivariantly projective if and only if $\Sigma$ is strongly polytopal. Now, let $\mathcal{X}_{\Sigma}$ be equivariantly projective and morphism
$\Phi$  be embedding which is induced by the rational map $\phi:\mathcal{X}_{A}  \longrightarrow  \mathbf{P}^{r}$
defined by $p \mapsto[z^{m_{0}},z^{m_{1}},\ldots,z^{m_{r}}],$ where $z^{m_{l}}(p)=p^{m_{l}}$ in case $p=(p_{1},p_{2},\ldots p_{n})$. Then, the rational map $\Phi(\mathcal{X}_{\Sigma§})$ is the set of common solutions of finitely many monomial equations
\begin{equation}
z^{\beta_{0}}_{i_{0}}z^{\beta_{1}}_{i_{1}}\cdots z^{\beta_{s}}_{i_{s}}=z^{\beta_{s+1}}_{i_{s+1}}z^{\beta_{s+2}}_{i_{s+2}}\cdots z^{\beta_{r}}_{i_{r}}
\end{equation}
which satisfy the following relationships
\begin{equation}
  \beta_{0}m_{0}+\beta_{1}m_{1}+\cdots +\beta_{s}m_{s}=\beta_{s+1}m_{s+1}+\beta_{s+2}m_{s+2}+\cdots +\beta_{r}m_{r}
\end{equation}
and
\begin{equation}
  \beta_{0}+\beta_{1}+\cdots +\beta_{s}=\beta_{s+1}+\beta_{s+2}+\cdots +\beta_{r}
,
\end{equation}
for all $\beta_{l}\in \mathcal{Z}_{\geq 0}$ and $l=0,1,\ldots, r$ \cite{Ewald}.
As we have seen for  multi-qubit  systems the separable states are given by the Segre embedding of $\mathbf{CP}^{1}\times\mathbf{CP}^{1}\times\cdots\times\mathbf{CP}^{1}$.
Now, for example, let $z_{1}=\alpha^{1}_{1}/\alpha^{1}_{0}, z_{2}=\alpha^{2}_{1}/\alpha^{2}_{0},\ldots, z_{m}=\alpha^{m}_{1}/\alpha^{m}_{0}$.
  Then we can cover $\mathbf{CP}^{1}\times\mathbf{CP}^{1}
  \times\cdots\times\mathbf{CP}^{1}$ by $2^{m}$ charts
\begin{eqnarray}
\nonumber &&
\mathbf{X}_{\check{\Delta}_{1}}=\{(z_{1},z_{2},\ldots,z_{m})\},\\\nonumber&&
~\mathbf{X}_{\check{\Delta}_{2}}=\{(z^{-1}_{1},z_{2},\ldots,z_{m})\},\\\nonumber&&
~~~~~~~~~~\vdots
\\\nonumber&&
\mathbf{X}_{\check{\Delta}_{2^{m}-1}}=\{(z_{1},z^{-1}_{2},\ldots,z^{-1}_{m})\},
\\\nonumber&&
\mathbf{X}_{\check{\Delta}_{2^{m}}}=\{(z^{-1}_{1},z^{-1}_{2},\ldots,z^{-1}_{m})\}
\end{eqnarray}
Let us consider the $m$
-hypercube $\Sigma$ centered at the origin with vertices $(\pm1,\ldots,\pm1)$. This gives the toric variety $\mathcal{X}_{\Sigma}=
\mathbf{CP}^{1}\times\mathbf{CP}^{1}\times\cdots\times\mathbf{CP}^{1}$.
Now, the map $\Phi(\mathcal{X}_{\Sigma})$ is  a set of the common
solutions of the following monomial equations
\begin{equation}
x^{\beta_{0}}_{i_{0}}x^{\beta_{1}}_{i_{1}}\cdots
x^{\beta_{2^{m-1}-1}}_{i_{2^{m-1}-1}}=x^{\beta_{2^{m-1}}}_{i_{2^{m-1}}}
\cdots x^{\beta_{2^{m}-1}}_{i_{2^{m}-1}}
\end{equation}
that gives  quadratic polynomials $\alpha_{k_{1}k_{2}\ldots
k_{m}}\alpha_{l_{1}l_{2}\ldots l_{m}} = \alpha_{k_{1}k_{2}\ldots
l_{j}\ldots k_{m}}\alpha_{l_{1}l_{2} \ldots k_{j}\ldots l_{m}}$ for
all $j=1,2,\ldots,m$ which  coincides with the Segre ideals.
Moreover, we have
\begin{equation}
\Phi(\mathcal{X}_{\Sigma})=\mathrm{Specm} \mathcal{C}[\alpha_{00\ldots
0},\alpha_{00\ldots 1},\ldots,\alpha_{11\ldots
1}]/\mathcal{I}(\mathcal{A}),
\end{equation}
where $\mathcal{I}(\mathcal{A})=\langle \alpha_{k_{1}k_{2}\ldots
k_{m}}\alpha_{l_{1}l_{2}\ldots l_{m}} - \alpha_{k_{1}k_{2}\ldots
l_{j}\ldots k_{m}}\alpha_{l_{1}l_{2} \ldots
k_{j}\ldots l_{m}}\rangle_{\forall j;k_{j},l_{j}=0,1}$.
This toric variety describe the space of separable states in a multi-qubit quantum systems.
In summary we have investigated the geometrical structures of quantum multi-qubit states based on the Segre variety  toric varieties. We showed that multi-qubit states can be characterized and visualized by embedding of toric variety in a complex projective space. The results are interesting in our voyage to the realm of quantum theory  and  a better understanding of the nature of  multipartite quantum systems.
\begin{flushleft}
\textbf{Acknowledgments:}  The  work was supported  by the Swedish Research Council (VR).
\end{flushleft}

\end{document}